\documentclass[prl,aps,twocolumn]{revtex4}

\usepackage{amsmath}
\usepackage{amssymb}
\usepackage{bm}

\newcommand{\beq}{\begin{equation}}
\newcommand{\eeq}{\end{equation}}
\newcommand{\bea}{\begin{eqnarray}}
\newcommand{\eea}{\end{eqnarray}}
\newcommand{\bei}{\begin{itemize}}
\newcommand{\eei}{\end{itemize}}

\def\vec#1{{\bf #1}}

\begin{document}

\title{Effect of Landau Level Mixing on Braiding Statistics}

\author{Steven H. Simon}
\affiliation{Bell Laboratories, Alcatel-Lucent, Murray Hill,
New Jersey 07974}

\begin{abstract}
We examine the effect of Landau level
mixing on the braiding statistics of quasiparticles of abelian and
nonabelian quantum Hall states.   While path dependent geometric
phases can perturb the abelian part of the statistics, we find that
the nonabelian properties remain unchanged to an accuracy that is
exponentially small in the distance between quasiparticles.
\end{abstract}
\date{\today}
\pacs{ } \maketitle

In 2+1 dimensions, quasiparticles may exist that obey nontrivial
braiding statistics\cite{RevModPhys}.  A well-known example occurs for the  $\nu=1/m$
Laughlin quantum Hall state, where
taking one quasiparticle adiabatically clockwise around another accumulates a
Berry's phase $e^{2 \pi i /m}$ which is independent of the details of the path, but depends only on its topology (i.e., that it went around the other particle).  Since the braiding statistics is comprised only of a phase, this case is known as ``abelian".   In more complicated
 ``nonabelian" quantum Hall states\cite{MooreRead}, the
ground state manifold can be expressed as a vector space, and
taking one particle around some others applies a unitary matrix to
this vector space, with the unitary matrix depending only on the topology of the path and not on its geometric details. This
independence from details can in principle allow highly robust
quantum information processing and is fundamental to the concept of
topological quantum computation\cite{RevModPhys}.

It it known, however, that certain effects can introduce corrections to the braiding statistics of abelian quantum Hall states\cite{Haldane,Sondhi,HannaLee}.  Most seriously, when Landau level (LL) mixing is taken into account, the long range Coulomb interaction can introduce corrections that are power law in the distance between quasiparticles.  If similar
corrections were to occur for the nonabelian braiding statistics of
nonabelian quantum Hall states, it would present a fundamental barrier
to the idea of topological quantum computation\cite{RevModPhys}. The
purpose of this paper is to show that this does not occur.  To be
precise, we assume that we are given a quantum Hall wavefunction,
fully in the lowest Landau level (LLL), which is in a well defined
topological phase (with all its associated topological properties)
and we then examine the effects of LL mixing.  We find, for LL mixing not too large, to accuracy exponentially small in the
distance between quasiparticles, only the abelian phases are
perturbed --- which then does not present a problem for the idea of topological quantum computation\cite{RevModPhys}.

At first glance, it might seem natural that nonabelian statistics
would be better protected from perturbations than abelian
statistics, as it is well known\cite{Ocneanu} that nonabelian
statistics can only occur in a discrete set of varieties (which cannot be perturbed continuously) whereas there is a continuum of possible
abelian statistics. However, this so-called Ocneanu rigidity\cite{Ocneanu} of
nonabelian statistics is not sufficient to show that the
nonabelian statistics are preserved under perturbation, as it
possible that the result of taking a quasiparticle around a loop may
end up being non-topological (i.e., may depend on the details of the
path and not just its topology), and hence the perturbed system may
not have a well defined braiding statistics at all. Indeed, this is
precisely what happens in the abelian case\cite{Sondhi} where the
correction (from long range interactions and LL mixing) to the phase
accumulated by taking one quasiparticle around another is dependent
on the distance between the quasiparticles.

We start by reviewing how one calculates geometric
phases and topological statistics associated with transporting
quasiparticles in the LLL\cite{Gurarie,RevModPhys}.  For nonabelian quantum Hall states, in the presence of quasiparticles, the
ground state is described by an arbitrary superposition of several wavefunctions $\Psi_i$, which we call ``blocks" (for the abelian case we have only a single block).   If
there are $N$ electron coordinates $z_i$ and $M$ quasiparticle
coordinates $w_\alpha$,  we write these blocks as
\begin{eqnarray}
\label{eq:blocks}
    \Psi_i  &=& \phi_i(z_1, \ldots, z_N; w_1, \ldots w_M) \\
    &\times& \exp \left[
-\frac{1}{4 \ell^2} \sum_{i=1}^N |z_i|^2 - \frac{e^*}{e}\frac{1}{4 \ell^2}  \sum_{\alpha=1}^M |w_\alpha|^2 \right]  \nonumber
    \end{eqnarray}
where
$\ell$ is the magnetic length, $e^*$ is the quasiparticle charge, and we choose to work in holomorphic gauge where $\phi$ is holomorphic in all of its arguments.  Although $\phi$ is single valued in the $z$ coordinates, it may be multiple valued in the $w$ coordinates (due to branch cuts), and this multiple valuedness is why there are multiple blocks $\phi_i$.  We assume that these
blocks are orthonormalized \begin{equation}
\label{eq:norm}
    \langle \Psi_i | \Psi_j \rangle = \delta_{ij}~~~~.
\end{equation}
If wavefunctions are generated as conformal blocks of a conformal field theory (CFT) with an appropriately chosen background charge, then the form of Eq.~\ref{eq:blocks} in holomorphic gauge appears
naturally\cite{MooreRead,RevModPhys}.  Orthonormality (Eq. \ref{eq:norm}) for conformal blocks is then often assumed\cite{Gurarie,RevModPhys} so long as quasiparticles are well separated, although this has only been proven for a few special cases\cite{ReadCooper,Wen}.

For nonabelian quantum Hall states, the effect of
adiabatically transporting a particle $w$  around a closed loop is to apply some unitary matrix  to the space of blocks $|\Psi_i\rangle \rightarrow U_{ij} |\Psi_j \rangle$.   This matrix is the product of the explicit monodromy matrix ${\cal M}$
(which gives the explicit change in the wavefunction due to branch
cuts when moving the coordinate $w$) and a Berry's matrix
${\cal B}$ given by\cite{RevModPhys,Wilczek}
\begin{equation}
\label{eq:berry}
    {\cal B} = {\cal P} \exp\left[ \int d\tau  \,  \left\langle
    \Psi_i \left| \frac{d}{d\tau} \right| \Psi_j \right\rangle \right]
\end{equation}
where $\tau$ indexes  the position of the quasiparticle, and
${\cal P}$ is the path ordering operator\cite{endnote5}.  Writing the quasiparticle position in holomorphic and antiholomorphic coordinates we have $d/d\tau = (d w/d\tau) \partial_w +
(d  \bar w / d\tau) \partial_{\bar w}$.   Noting that  $\phi_i$ is holomorphic,  $\partial_{\bar w}$ applied to $|\Psi_j\rangle$ acts only
on the gaussian factors to give $\langle \Psi_i | \partial_{\bar w} | \Psi_j\rangle = -\delta_{ij} (e^*/e) (w/4 \ell^2)$.  Similarly, using Eq.~\ref{eq:norm} we have $\partial_w \langle \Psi_i | \Psi_j \rangle = 0$ so we can replace $\langle \Psi_i | \partial_{w} | \Psi_j\rangle$ by $-(\partial_{w} \langle \Psi_i |)   | \Psi_j\rangle$ so that $\partial_w$ acts
only on the gaussians. We then easily obtain the abelian Berry's matrix ${\cal B} = \delta_{ij} e^{i \gamma}$ where
\begin{equation}
\label{eq:area}
    \gamma  =  (A/\ell^2)(e^*/e)
\end{equation}
where $A$ is the area surrounded by the path.  The fact that this is proportional to the area enclosed means that it is a {\it geometric} phase, and is therefore not considered part of the braiding statistics. Thus, the braiding statistics come entirely from the explicit monodromy of the wavefunction in holomorphic gauge.  (For wavefunctions generated by CFT, the nonabelian statistics is thus the monodromy of the conformal blocks, assuming Eq.~\ref{eq:norm}).

The point of this paper is to extend this calculation to the case
where  LL mixing is included as a
perturbation.  Our starting point is to {\it assume} that the
unperturbed state (fully in the LLL) is indeed a topological state of matter with all of its associated properties\cite{RevModPhys}.    In particular we
will assume  (a) that if quasiparticles remain well separated then  (to accuracy exponentially small in the
separation between quasiparticles for large enough separations) blocks cannot be distinguished from each other by any topologically local measurement\cite{endnote0}.  In addition, we note (b) that since the system is in a gapped phase, any sufficiently small change in the LLL Hamiltonian will not alter the topological phase of the system in the LLL, and in particular will not destroy property (a).   This is consistent with the fact that for many quantum Hall states there is a broad range of LLL Hamiltonians that produce essentially the exact same wavefunctions.

Let us now be a bit more precise about assumption (a).
Given an operator $\hat O$ that acts nontrivially on electrons only
in a region ${\cal R}$, we say that the operator is ``topologically
local" if the region ${\cal R}$ consists of
disconnected (well separated) regions each of which surrounds only a single quasiparticle or no quasiparticles.   Assumption (a) is equivalent to the statement that for any topologically local $\hat O$ we must have $\langle \Psi_i | \hat O | \Psi_j
\rangle = C \delta_{ij}
$ for some $C$ (which may be a function of
the $w$ coordinates). This assumption is crucial to the idea
of using quantum Hall states for topological quantum
computation, as the inability of any local operator to
distinguish blocks is precisely the property that makes the space of
blocks highly immune to decoherence.  We emphasize again that this statement has not
been proven in general, but is frequently assumed to be true.  Note that if $\hat O$ is any sum of topologically local operators, then  $\langle \Psi_i | \hat O | \Psi_j
\rangle = C \delta_{ij}$ remains true, and we think of this as also being topologically local.

To understand the effects of LL mixing, we first define the interaction operator $V = \int \vec d\vec q  \, v(|\vec q|) \,
 \rho(\vec q) \rho(-\vec q)   $, where $\rho(\vec q)$ is the density
operator at wavevector $\vec q$ and $v(q) =2 \pi e^2 \exp(-q \ell_v) /( \epsilon q )$ is the Coulomb interaction (with $\epsilon$ the dielectric constant) which we have cut
off at some finite short length scale $\ell_v$ to avoid
singularities.  (This cutoff may be provided in a physical system
by finite well width, for example).  It is convenient to work with a coherent state basis\cite{Kivelson} for single particle wavefunctions which we write as $\varphi_{\vec R,n}(\vec r) = |\vec R, n \rangle $ where $n$ is the  LL index and $\vec R$ is the guiding center coordinate.  Although these coherent states are non-orthogonal, we still have the completeness relation ${\bf 1} =  (2 \pi \ell^2)^{-1} \sum_n \int
\vec d\vec R \, |\vec R, n\rangle \langle \vec R, n|$. We define annihilation operators for an electron
in such a coherent state $c_{\vec R,n} = \int \vec d\vec r \,
\varphi_{\vec R,n}(\vec r) \psi(\vec r)$ and using completeness obtain $\psi(\vec r)
= (2 \pi \ell^2)^{-1} \sum_n \int \vec d\vec R \, \varphi_{\vec R,n}(\vec
r)c_{\vec R,n} $, where $\psi$ is the usual electron annihilation
operator.  The density operator
can then be written as
\begin{equation}
    \rho(\vec q) = \mbox{$\sum_{n,n'}$} \int \! \frac{\vec d\vec R \, \vec d \vec R'}{(2\pi \ell^2)^2} \,  \langle \vec R n |
    e ^{i \vec q \cdot \vec r} | \vec R' n' \rangle   \, c^{\dagger}_{\vec R
    n} c^{\phantom{\dagger}}_{\vec R' n'}
\end{equation}
A key observation here is that the matrix element is local:
$|\langle \vec R n | e ^{i \vec q \cdot \vec r} | \vec R' n'
\rangle|$ reaches a maximum of order unity or less
at $\vec R - \vec R' = \hat z \times \vec q \ell^2$ (where $\hat z$
points normal to the plane) and decays exponentially for all
$|\vec R - \vec R'| > \vec q \ell^2 $. Since we have cut off the interaction at
$|\vec q| \sim 1/\ell_v$ this means that in a single application of
the interaction $V \sim \rho(\vec q) \rho(-\vec q)$, no electron guiding center coordinate can be moved more than a distance of order $\ell^2/\ell_v$.  Note that even though $V$ is long ranged, it is a sum (or integral) of topologically local operators.

We now define an operator $P_0$ to be projection of all electrons to the LLL, and define $\hat V = V -  P_0 V P_0$ to be the interaction, with the LLL part removed (i.e., $\hat V$ is the part of the interaction that we want to treat as a perturbation). The unperturbed Hamiltonian will be $H_0 = K + P_0 V P_0$ with $K$ the kinetic energy, and the full Hamiltonian is then $H = K + V = H_0 + \hat V$.   Given an eigenstate $|\Psi\rangle$ of the full Hamiltonian, we define the LLL piece, $|\Psi^0\rangle = P_0 |\Psi \rangle$.  It turns out to be convenient to work with a normalization where $\langle \Psi^0 |  \Psi^0 \rangle=1$.   We thus write $|\Psi^u \rangle =  |\Psi^0 \rangle + |\Psi^\perp\rangle$, where the superscript $u$ means ``unnormalized" and we construct the
normalized wavefunction $|\Psi \rangle= {\cal Z}^{1/2} |\Psi^u\rangle$ with ${\cal Z}^{-1} = \langle \Psi^u | \Psi^u \rangle$.

We write the Schroedinger equation as $|\Psi^u \rangle = (E-H_0)^{-1} \hat V |\Psi^u \rangle$ and then
project with $1 - P_0$ to obtain $|\Psi^\perp\rangle = \hat G(E) \hat V ( |\Psi^0 \rangle + |\Psi^\perp\rangle)$ where we have defined  $\hat G(E) = (1 - P_0) (E-H_0)^{-1} = \sum^{\phantom k}_{k_{\phantom k}} |k\rangle\langle k| / (E - \varepsilon_k)$ where the sum is over only states $k$ (with unperturbed energy $\varepsilon_k$) which are not fully in the LLL.   Note that the energy denominator is (at least) of order the cyclotron energy $\hbar \omega_c$.   We then obtain the full wavefunction $|\Psi^u \rangle = \hat X(E)  |\Psi^0 \rangle$ where  $\hat X(E) = \sum_{k=0}^\infty (\hat G(E) \hat V)^k$.    We have thus written a perturbation expansion for the full wavefunction $|\Psi^u\rangle$ in terms of its LLL component $|\Psi^0\rangle$.

We now find an effective theory for wavefunctions fully in the LLL.   We can rewrite the Schroedinger equation now as $H \hat X  |\Psi^0 \rangle = E \hat X   |\Psi^0 \rangle$ which, projected to the LLL,  can then be recast as
$H_{eff}| \Psi^0 \rangle = E  | \Psi^0 \rangle$, with the effective Hamiltonian $H_{eff} =  H_0  + \delta H_0 $ where $\delta H_0 = P_0 \hat V \hat X P_0 = P_0 [ \hat V \hat G \hat V  + \hat V \hat G \hat V \hat G \hat V + \ldots] P_0$.  Here $\delta H_0$ is a LLL interaction that includes all of the effect of  integrating out higher LL's exactly.  The perturbed energy is thus given by  $E=  \varepsilon_0 +  \langle \Psi^0 | \delta H_0 |  \Psi^0 \rangle$ where $\varepsilon_0 = \langle \Psi^0 | H_0 |  \Psi^0 \rangle$ is the unperturbed energy.

Our initial assumption is that the unperturbed system ($H_0$ in the LLL) is in a gapped topological phase.   As mentioned above in (b) since the state is assumed to be gapped, perturbing the Hamiltonian slightly with the LLL interaction $\delta H_0$ cannot change the topological phase of matter within the LLL.   Thus, we assume that the system defined by $H_0 + \delta H_0$ fully in the LLL is also in this topological phase, and in particular satisfies assumption (a) above.  We denote the blocks of this perturbed LLL systems as $|\Psi^0_i\rangle$.  From these blocks, we can use our above perturbation theory to generate the corresponding full wavefunctions $|\Psi_i\rangle$ as described above.  What we will show below is that the nonabelian part of the braiding statistics of the blocks $|\Psi_i \rangle$ of the full multi-LL system  must be the same as that of the LLL blocks $|\Psi_i^0 \rangle$.

%
%

First, let us re-examine Eq.~\ref{eq:norm} for the blocks of the full system.  We define $[\hat X^\dagger \hat X]_n$ to be the operator $\hat X^\dagger \hat X$ expanded to $n^{th}$ order in perturbation theory.   (Each order in perturbation theory is a factor of $\hat G \hat V $ and is order $E_{coulomb}/\hbar \omega_c  = e^2/(\epsilon \ell \hbar \omega_c)$ ).   As emphasized above, each application of $\hat V$ can only move an electron a distance of order $\ell^2/\ell_v$ (and application of $\hat G$ does not move anything at all).   Thus, the operator $[\hat X^\dagger \hat X]_n$ can only move particles a maximum distance of $n \ell^2/\ell_v$, and is therefore topologically local so long as quasiparticles are a distance $d \gtrsim n \ell^2/\ell_v$ apart.  Since the LLL wavefunction $|\Psi^0_i\rangle$ is in a topological phase, assumption (a) tells us that $\langle \Psi_i^0 | [X^\dagger X]_n |  \Psi^0_j \rangle = C \delta_{ij}$ with accuracy increasing exponentially with $d$ for  $d \gtrsim n \ell^2/\ell_v$.    So long as our perturbation theory converges\cite{endnote4}, then for sufficiently large $n$ the difference between the matrix element of  $[\hat X^\dagger \hat X]_n$ and $\hat X^\dagger \hat X$ will decrease exponentially quickly with increasing $n$.  Thus, for sufficiently large $d$, we have $\langle \Psi_i^0 | \hat X^\dagger \hat X |  \Psi^0_j \rangle = C \delta_{ij}$ with accuracy increasing exponentially with $d$.  Thus, to this exponential precision, Eq.~\ref{eq:norm} is obeyed by the full wavefunctions $|\Psi_i\rangle$.  Furthermore the normalization constant $\langle \Psi_i^u | \Psi_i^u \rangle ={\cal Z}^{-1}$ is independent of the block $i$.  Similarly, using locality of $\delta H_0$, we show that the energy $E$ is also independent of the block.

We are now prepared to calculate the braiding statistics of the system with LL mixing. As above, we first calculate the Berry's matrix Eq.~\ref{eq:berry}.    Here we use the wavefunctions $|\Psi_j\rangle = {\cal Z}^{1/2}|\Psi^u_j \rangle = {\cal Z}^{1/2} \hat X |\Psi^0_j \rangle$.   We must now be cautious because $\hat X$ and  ${\cal Z}$ depend not only on the quasiparticle position, but also on the energy $E$.  We thus write
$d/d\tau = (d w/d\tau) \partial_{w|E} +  (d\bar w/d \tau) \partial_{\bar w|E} + (d E/d\tau)  \partial_{E|w,\bar w}$ where we use derivative notation where parameters listed after the $|$ in the subscripts are held constant.   Fortunately,  as shown in the appendix, $\langle \Psi_i | \partial_{E|w,\bar w} | \Psi_j\rangle = 0$, so we can drop this term from $d/d\tau$.   We now proceed as above in the unperturbed LLL case.  Application of $\partial_{\bar w|E}$ to  $|\Psi_j\rangle$ now acts only on the gaussian factors of $|\Psi_j^0\rangle$ and on ${\cal Z}$.  We obtain
$\langle \Psi_i | \partial_{\bar w|E} | \Psi_j\rangle = \delta_{ij} [-(e^*/e) (w/4 \ell^2) + (1/2) \partial_{\bar w|E} \log {\cal Z}]$.  As above we use orthonormality (Eq.~\ref{eq:norm}) to move $\partial_{w|E}$  to act on $\langle \Psi_i|$ giving a similar result.     We again end up with an abelian Berry's matrix with phase
$\gamma + \delta \gamma$ where $\gamma$ is as given above in Eq.~\ref{eq:area} and $\delta \gamma = - {\rm Im} \oint dw \, \partial_{w|E} \log [{\cal Z}^{-1}]$. Crucially, there is no nonabelian contribution from the explicit Berry's matrix, although the abelian contribution has a correction. Thus the nonabelian braiding statistics are again entirely given by the explicit monodromy.  However, since both the normalization ${\cal Z}^{1/2}$ and the energy $E$ are independent of the block, and (by linearity) the monodromy of $\hat X |\Psi^0_i\rangle$ is obviously the same as that of $|\Psi^0_i\rangle$, we find that the monodromy of the full wavefunction $|\Psi_i\rangle$ is the same as that of its LLL component  $|\Psi_i^0\rangle$.    This is the main technical result of this paper: the nonabelian part of the braiding statistics is independent of projection to the LLL
(subject to the same condition that the quasiparticles are far enough apart and corrections are exponentially small in this distance).  Note that our perturbative analysis relies on having a cutoff for the Coulomb interaction at some finite distance $\ell_v$  (which is indeed true in experimental systems).   One can ask if the result remains true without this cutoff.   While  our derivation cannot handle this case rigorously, it seems unlikely such details of the interaction would be important.

We note in passing that we can use similar techniques to study perturbations of systems completely in the LLL.  For example, given a solvable Hamiltonian $H_0$ with a topological ground state in the LLL (such as the special interaction that gives the Read-Rezayi wavefunction as its exact ground state\cite{ReadRezayi}), and  given a perturbation interaction $\hat V$ in the LLL, we can use an analogous argument to study $H_0 + \hat V$ where the perturbation parameter is now $\hat V/\Delta$ with $\Delta$ the excitation gap of $H_0$.   Indeed, assuming (a) holds for blocks of $H_0$, we can show that the blocks of $H_0 + \hat V$ also satisfy (a) so long as $\hat V$ is appropriately topologically local.   This can be used as a rough ``proof" of assumption (b) above.

We now return examine the abelian phase $\gamma + \delta \gamma$ calculated above.
Since ${\cal Z}^{-1} =  \langle \Psi^0 | \hat X^\dagger \hat X |  \Psi^0 \rangle = 1 + \langle \Psi^0 | \hat V \hat G^2 \hat V  |  \Psi^0 \rangle + \ldots$, the correction $\delta \gamma$ is order $[E_{coulomb} / \hbar \omega_c]^2$.
This result apparently contradicts prior work\cite{Sondhi,HannaLee} which suggest corrections at first order.   However, there is a subtlety that we have not yet considered:  in the unperturbed wavefunction $\Psi^0_i$, the parameter $w$ represents the position of the quasiparticle (i.e., its complex coordinate is $z=w$).   However, once the wavefunction is perturbed, the quasiparticle's complex coordinate will generally move to some perturbed position $z = w + \delta w$. Neglecting the second order correction $\delta \gamma$, the geometric phase associated with moving the  {\it parameter} $w$ around a given path remains given by $\gamma$ in Eq.~\ref{eq:area}   where $A$ is the area enclosed by the path {\it in parameter space}.  However because the quasiparticle's position is perturbed, the {\it physical} path of the quasiparticle is not the same as the path in parameter space and does not enclose an area $A$.


We now estimate the magnitude of this effect.  Let us imagine a quasiparticle at the origin as the source of an unscreened long range potential $U(r) = e e^*/(\epsilon r)$.   We first solve the problem of noninteracting electrons exposed to this potential.  In symmetric gauge, we write the single particle wavefunction with angular momentum $m$ as $\psi_m = e^{i m \theta} y(r) r^{-1/2}$ and we obtain an effective one dimensional Schroedinger equation $-(\hbar \omega_c/2) d^2y/dr^2  +  u(r) y = E y$ where the effective potential is $u(r) = U(r) + (\hbar \omega_c/8) [-4 m + (4 m^2-1) (\ell/r)^2 + (r/\ell)^2] $.   Solving this perturbed Schroedinger equation mixes LL's of the unperturbed basis.   The resulting (perturbed) ground state wavefunction is very peaked near the minimum of the effective potential, which in the long distance and weak interaction limit is at position $r_{max} \approx \sqrt{2 m} \, \ell + \delta r$ where $\delta r = \ell \, [e^*/(2 e m)] [(e^2/\epsilon \ell)/(\hbar \omega_c)]$.  This shift in position of each orbital (assuming no other change to the multiparticle wavefunction) gives precisely the $1/r^3$ corrections to the density profile of the quasiparticle found in Refs.~\cite{HannaLee,Sondhi}.  As a result, any other quasiparticle at distance $r$ from the one at the origin will accordingly have its position shifted by $\delta r$,  although its position will still be labeled by the original parameter $w$.  To return the quasiparticle to position $z=w$, the parameter $w$ needs to be changed by roughly $-\delta r$.  Taking this quasiparticle around the one at the origin then gives an extra abelian geometric phase $-(2 \pi r (\delta r)/\ell^2)(e^*/e) =    -(2 \pi \ell/r) (e^*/e)^2 [(e^2/\epsilon \ell)/(\hbar \omega_c)]$ in comparison to the case without the long range interaction and LL mixing.

It has also been pointed out that there is yet another source of correction to the abelian statistics due electro-dynamical effects\cite{Haldane,HannaLee}.  Here, the currents in the quantum Hall layer generate a very small amount of actual physical flux, and this physical flux binding slightly changes both the abelian statistics and the total charge of the quasiparticle\cite{Haldane,HannaLee}.  However, this mechanism again has no way to effect the nonabelian braiding statistics.

To summarize, we have demonstrated that, to exponential accuracy with increasing distance between quasiparticles, there are no corrections from long range interactions and LL mixing to the nonabelian statistics of quantum Hall states.   This result is crucial to the idea of using such quantum Hall states for quantum information processing\cite{RevModPhys}.   We have also examined perturbations to the abelian phase from LL mixing and find several contributions, the leading term of which agrees with previous works\cite{Sondhi,HannaLee}.  While these calculations have all been done for LLL wavefunctions, it is trivial to generalize to the higher LL's, where we suspect that we may have seen actual nonabelian states of matter experimentally\cite{RevModPhys}.


The author acknowledges conversations with N.~Bonesteel, V.~Gurarie, M.~Freedman, J.~Slingerland, and F.D.M.~Haldane. 

{\it Appendix:} Throughout this appendix we fix quasiparticle positions.    As in the main text we use locality of $\hat X^\dagger \partial_E \hat X$ to show $\langle \Psi_i | \partial_E | \Psi_j \rangle= \langle \Psi_i^0 | \hat X^\dagger \partial_E \hat X| \Psi_j^0 \rangle= C \delta_{ij}$.   Since  $|\Psi_i \rangle = {\cal Z}^{1/2} \hat X|\Psi_i^0\rangle$ is normalized,  $\langle \Psi_i | \partial_E | \Psi_i\rangle$ can only be nonzero if it is imaginary.  Using  $\partial_E \hat G = -\hat G^2$ we show $\partial_E \hat X = -\hat X \hat G \hat X + \hat X \hat G$ so that $\langle \Psi_i^0 | \hat X^\dagger \partial_E \hat X |  \Psi_i^0\rangle =  -\langle \Psi_i^0 | \hat X^\dagger \hat X \hat G \hat X |  \Psi_i^0\rangle$  since $\hat G |\Psi^0_i \rangle =0$.  However, since  $\hat X \hat G = \hat G \hat X^\dagger$ and $\hat G$ is hermitian, we have  $\hat X^\dagger \hat X \hat G \hat X$ hermitian.   Thus the matrix element is real  so $\langle \Psi_i | \partial_E | \Psi_j\rangle=0$.


\begin{thebibliography}{14}
\bibitem{endnote5}   Only the product  ${\cal B M}$ is gauge invariant\cite{Gurarie,RevModPhys}.
\bibitem{endnote4} We expect that as long as $H_0$ is not on a phase boundary (such as the fermi liquid at zero interaction strength) the perturbation series will be convergent rather than asymptotic.
\end{thebibliography}
\end{document}